\documentstyle[12pt,epsf]{article}

\begin{document}
\begin{flushright}{UT-02-33}
\end{flushright}

\sloppy
\sloppy
\sloppy

\vskip 0.5 truecm

\begin{center}
{\large{\bf  Path Integral for Separable Hamiltonians of 
Liouville-type }}
\end{center}
\vskip .5 truecm
\centerline{\bf Kazuo Fujikawa}
\vskip .4 truecm
\centerline {\it Department of Physics,University of Tokyo}
\centerline {\it Bunkyo-ku,Tokyo 113,Japan}
\vskip 0.5 truecm

\makeatletter
\@addtoreset{equation}{section}
\def\theequation{\thesection.\arabic{equation}}
\makeatother

\vskip 0.5 truecm

\begin{abstract}
A general path integral analysis of the separable Hamiltonian of 
Liouville-type is reviewed. The 
basic dynamical principle used is the Jacobi's principle of 
least action for given energy which is reparametrization 
invariant, and thus the gauge freedom naturally appears.
The choice of gauge in path integral corresponds to the  
separation of variables in operator formalism.
The  gauge independence and the operator ordering are closely 
related. The path integral in this formulation 
 sums over orbits in space instead of space-time.
An  exact path integral of the Green's function for the hydrogen 
atom in parabolic coordinates is ilustrated as an example, 
which is also interpreted as one-dimensional
quantum gravity with a  quantized cosmological constant.
\end{abstract}

\newpage
\section{Introduction}
 In 1979,  Duru and Kleinert[1] showed an elegant  
path integral method to evaluate the Green's function for the 
hydrogen atom exactly. Two basic 
ingredients in their method are the use of a re-scaled time 
variable and the  so called  Kustaanheimo-Stiefel 
transformation[2] which reveals the O(4)
symmetry explicitly in the coordinate space. The 
physical meaning of the ``re-scaled time variable'' however 
remained somewhat unclear[3]. I have studied this issue by 
recognizing the procedure in Ref.[1] as a special case of the 
general treatment of classically separable Hamiltonian of 
Liouville-type. The basic dynamical principle involved is 
then identified as the Jacobi's principle of least action for
 given energy[4]. 

The path integral on the basis of the Jacobi's principle of 
least action is basically static and analogous to geometrical 
optics, and one deals with a 
sum over orbits in {\em space} instead of space-time.   
Another characteristics of the Jacobi's
principle of least action is  that it is reparametrization 
invariant. The general technique of gauge 
theory is thus applicable to the evaluation of path integral, 
and  a suitable choice of gauge  simplifies the problem such as 
the hydrogen atom.
A simple trick in parabolic coordinates, which was used before 
in a different context by Ravndal and Toyoda[5], renders 
the hydrogen atom Hamitonian a separable form of Liouville-type.

The hydrogen atom path integral is also understood as a 
one-dimensional quantum gravity; an interesting aspect of this 
picture is that the cosmological constant is {\em quantized}[6].

\section{Separable Hamiltonian of Liouville-type}

We start with a separable Hamiltonian 
\begin{equation}
H= \frac{1}{V_{1}(q_{1}) + V_{2}(q_{2})}\{\frac{1}{2m}(p_{1}^{2} + 
p_{2}^{2}) + U_{1}(q_{1}) + U_{2}(q_{2})\}
\end{equation}
where the variables change over $\infty > q_{1}, q_{2} >- \infty$. A 
general Hamiltonian of Liouville-type is given by 
\begin{equation}
H= \frac{1}{V_{1}(Q_{1}) + V_{2}(Q_{2})}\{\frac{1}{2m W_{1}(Q_{1})}P_{1}
^{2} + 
\frac{1}{2m W_{2}(Q_{2})}P_{2}^{2} + U_{1}(Q_{1}) + U_{2}(Q_{2})\}
\end{equation}
but after a canonical transformation 
\begin{eqnarray}
\frac{1}{\sqrt{W_{1}(Q_{1})}}P_{1}= p_{1} &,& 
\int_{0}^{Q_{1}}\sqrt{W_{1}(Q)}dQ 
= q_{1}\nonumber\\
\frac{1}{\sqrt{W_{2}(Q_{2})}}P_{2}= p_{2} &,& 
\int_{0}^{Q_{2}}\sqrt{W_{2}(Q)}dQ 
= q_{2} 
\end{eqnarray}
and a suitable redefinition of $V$ and $U$, we can write the 
 Hamiltonian in the form of (2.1).

We may then solve the Schroedinger problem
\begin{equation}
E\psi = \frac{1}{V_{1}(q_{1}) + 
V_{2}(q_{2})}\{\frac{1}{2m}(\hat{p}_{1}^{2} + 
\hat{p}_{2}^{2}) + U_{1}(q_{1}) + U_{2}(q_{2})\}\psi 
\end{equation}
where
\begin{equation}
\hat{p}_{l} = -i\hbar\frac{\partial}{\partial q_{l}}
\end{equation}
for $l=1, 2$,and the volume element $dV$, which renders the Hamiltonian 
$H$
in (2.4) hermitian, is given by
\begin{equation}
dV = (V_{1}(q_{1}) + V_{2}(q_{2}))dq_{1}dq_{2}
\end{equation}
The classical Hamiltonian (2.1) does not completely specify the operator 
ordering in (2.4), and the simplest ordering is adopted here. A precise 
operator ordering needs to be fixed depending on each 
explicit example.

One may rewrite the above Schroedinger equation (2.4) as 
\begin{equation}
\hat{H}_{T} \psi = 0
\end{equation}
with a {\em total Hamiltonian} defined by a specific gauge condition, 
\begin{equation}
\hat{H}_{T} = \frac{1}{2m}(\hat{p}_{1}^{2} + \hat{p}_{2}^{2}) + 
U_{1}(q_{1}) + U_{2}(q_{2}) - E(V_{1}(q_{1}) + V_{2}(q_{2}))
\end{equation}
The meaning of a total Hamiltonian is clarified later.
A general procedure to deal with a completely separated operator 
$\hat{H}_{T}$ is to 
consider an evolution operator for a parameter $\tau$ defined by
\begin{eqnarray}
\langle q_{1b},q_{2b}| e^{- i\hat{H}_{T}\tau/\hbar} | q_{1a}, 
q_{2a}\rangle
&=& \langle q_{1b}| \exp{[- (i/\hbar)(\frac{1}{2m}\hat{p}_{1}^{2}  + 
U_{1}(q_{1}) 
          - EV_{1}(q_{1}))\tau]} | q_{1a}\rangle\nonumber\\
&\times&\langle q_{2b}| \exp{[- (i/\hbar)(\frac{1}{2m}\hat{p}_{2}^{2}  + 
U_{2}(q_{2}) 
          - EV_{2}(q_{2}))\tau]} | q_{2a}\rangle\nonumber\\
&=& \int{\cal D}q_{1}{\cal D}p_{1}e^{(i/\hbar)\int_{0}^{\tau}\{ 
p_{1}\dot{q_{1}} 
- (\frac{1}{2m}p_{1}^{2}  + U_{1}(q_{1}) - 
EV_{1}(q_{1}))\}d\tau}\nonumber\\
&\times&\int{\cal D}q_{2}{\cal D}p_{2}e^{(i/\hbar)\int_{0}^{\tau}\{ 
p_{2}\dot{q_{2}} - (\frac{1}{2m}p_{2}^{2}  + U_{2}(q_{2}) - 
EV_{2}(q_{2}))\}d\tau}
\end{eqnarray}
The parameter $\tau$ is arbitrary , and by integrating over $\tau$ from 
$0$ to $\infty$ one obtains a physically meaningful quantity
\begin{eqnarray}
&&\langle q_{1b},q_{2b}|\frac{\hbar}{\hat{H}_{T}} | q_{1a}, 
q_{2a}\rangle_{semi-classical}
\nonumber\\
&=&i \int_{0}^{\infty}d\tau \frac{1}{\sqrt{2\pi i\hbar (\partial 
q_{1}(\tau)/
\partial p_{1}(0))_{q_{1a}}}}\frac{1}{\sqrt{2\pi i\hbar (\partial 
q_{2}(\tau)/
\partial p_{2}(0))_{q_{2a}}}}\nonumber\\
&\times&\exp{\{(i/\hbar)S_{cl}(q_{1b},q_{1a},\tau) + 
(i/\hbar)S_{cl}(q_{2b},q_{2a},\tau)\}}
\end{eqnarray} 
where we wrote the result of a semi-classical approximation for the path 
integral[7], though in certain cases one may be able to perform an exact 
path integral in (2.9). The pre-factor  in (2.10) is written in terms of  
classical 
paths , for example,
\begin{equation}
q_{1cl}(\tau) = q_{1}(\tau; q_{1a}, p_{1}(0))
\end{equation}
Namely, the classical paths dictated by the total Hamiltonian 
$\hat{H}_{T}$ are
expressed as functions of the initial positions and momenta. On the 
other hand,
the classical
action $S_{cl}$ is expressed as a function of the initial position, 
final
position and elapsed ``time'' $\tau$ by eliminating $p_{1}(0)$ 
dependence;for 
example, 
\begin{equation}
S_{cl}(q_{1b},q_{1a},\tau) = \int_{0}^{\tau}\{ p_{1}\dot{q_{1}} 
- (\frac{1}{2m}p_{1}^{2}  + U_{1}(q_{1}) - EV_{1}(q_{1}))\}_{cl}d\tau
\end{equation}
with $q_{1}(\tau) = q_{1b}$. If one solves the Hamilton-Jacobi equation 
in the 
form
\begin{equation}
S(q_{1b}, q_{1a}; \tau) = - A\tau + S(q_{1b}, q_{1a}; A )
\end{equation}
one treats $A$ as a dynamical variable and regards the above equation as 
a Legendre transformation defined by
\begin{eqnarray}
\frac{\partial S(q_{1b}, q_{1a}; \tau)}{\partial \tau} &=& -A\nonumber\\
\frac{\partial S(q_{1b}, q_{1a}; A )}{\partial A} & = & \tau
\end{eqnarray}
The variable $A$ is then eliminated.  This may be regarded as a 
classical analogue of uncertainty
relation; if one specifies $\tau$, the conjugate variable $A$ becomes
implicit. It is well-known that the semi-classical 
approximation (2.10) is exact for a 
quadratic system. 

We next note the relation for the quantity defined in the left-hand side 
of (2.10) 
\begin{eqnarray}
&&\langle q_{1b},q_{2b}|\frac{1}{\hat{H}_{T}} | q_{1a}, q_{2a}\rangle
\nonumber\\
&=&\langle q_{1b},q_{2b}|\frac{1}{(\frac{1}{\hat{V}_{1}(q_{1}) +
\hat{V}_{2}(q_{2})})\hat{H}_{T}} | q_{1a}, 
q_{2a}\rangle\frac{1}{V_{1}(q_{1a}) + V_{2}(q_{2a})}\nonumber\\
&=&\langle q_{1b},q_{2b}|\frac{1}{\hat{H} - E} | q_{1a}, 
q_{2a}\rangle\frac{1}{V_{1}(q_{1a}) + V_{2}(q_{2a})}\\
&=&\frac{1}{H( q_{1b},\frac{\hbar}{i}\frac{\partial}{\partial q_{1b}}, 
..) - E}\{\frac{1}{\sqrt{V_{1}(q_{1b}) + V_{2}(q_{2b})}}\langle 
q_{1b},q_{2b}| q_{1a}, q_{2a}\rangle\frac{1}{\sqrt{V_{1}(q_{1a}) + V_{2}
(q_{2a})}}\}\nonumber
\end{eqnarray}
by recalling $(\hat{A}\hat{B})^{-1} =\hat{B}^{-1}\hat{A}^{-1}$. The 
state vectors in these 
relations are defined for the volume element $dq_{1}dq_{2}$ as
\begin{equation}
\int dq_{1}dq_{2} |q_{1},q_{2}\rangle \langle q_{1},q_{2}| = 1
\end{equation}
Note that the definition of the $\delta$-function in $\langle 
q_{1}^{\prime},q_{2}^{\prime}|q_{1},q_{2}\rangle = \delta 
(q_{1}^{\prime}- q_{1})\delta (q_{2}^{\prime}- q_{2})$ depends on the 
choice of the volume elememt in (2.16) and thus on the choice of 
$H_{T}$.
The last expression in (2.15) is thus correctly defined for the original
Hamiltonian $H$ and the original state $\psi$ in (2.4)  with the volume 
element $dV$
in (2.6), since we have the completeness relation from (2.16)
\begin{equation}
\int  |q_{1},q_{2}\rangle \frac{dV}{V_{1}(q_{1}) + V_{2}(q_{2})}\langle 
q_{1},q_{2}| = 1 
\end{equation}
The left-hand side of (2.15) thus defines the correct Green's function 
for the original operator $(\hat{H} - E)^{-1}$ by noting the symmetry in 
$q_{a}$ and $q_{b}$.  One can then define the conventional evolution 
operator by 
\begin{eqnarray}
&&\langle q_{1b},q_{2b}|e^{ -i\hat{H}(t_{b} - t_{a})/\hbar} | q_{1a}, 
q_{2a}\rangle _{conv}\nonumber\\
&& = \frac{1}{2\pi i\hbar}\int_{-\infty}^{\infty} dE e^{-i E(t_{b} - 
t_{a})/\hbar}\langle q_{1b},q_{2b}|\frac{\hbar}{\hat{H}-i\epsilon - E} | 
q_{1a}, q_{2a}\rangle\frac{1}{V_{1}(q_{1a}) + V_{2}(q_{2a})}
\end{eqnarray}
where $\epsilon$ is an infinitesimal positive number. 
The total Hamiltonian changes for a different choice of gauge condition 
in the 
Jacobi's principle of least action to be explained below. Consequently, 
the 
volume element, which renders $H_{T}$ hermitian, generally depends on 
the 
choice of gauge. In the present case, one has the relation 
\begin{eqnarray}
\langle q_{1b},q_{2b} | q_{1a}, q_{2a}\rangle_{conv} &=&\langle 
q_{1b},q_{2b} | q_{1a}, q_{2a}\rangle\frac{1}{V_{1}(q_{1a}) + 
V_{2}(q_{2a})}\nonumber\\
 &=& \frac{1}{\sqrt{V_{1}(q_{1b}) + V_{2}(q_{2b})}}\langle q_{1b},q_{2b}
|q_{1a}, q_{2a}\rangle\frac{1}{\sqrt{V_{1}(q_{1a}) + V_{2}(q_{2a})}} 
\nonumber
\end{eqnarray}

\section{Jacobi's Principle of Least Action}

The meaning of the total Hamiltonian $H_{T}$ (2.8) becomes transparent 
if one 
starts with the Jacobi's principle of least action for a given $E$
\begin{eqnarray}
S& =& \int_{0}^{\tau} d\tau L = \int_{0}^{\tau}d\tau 
\sqrt{2m[E(V_{1}(q_{1}) + V_{2}(q_{2})) - (U_{1}(q_{1}) + U_{2}(q_{2}))]
(\dot{q}_{1}^{2} + \dot{q}_{2}^{2})}\nonumber\\
&=& \int \sqrt{2m[E(V_{1}(q_{1}) + V_{2}(q_{2})) - (U_{1}(q_{1}) + U_{2}
(q_{2}))][(dq_{1})^{2}+ (dq_{2})^{2}]}
\end{eqnarray}
which is reparametrization invariant. One then defines the momenta 
conjugate to coordinates
\begin{equation}
p_{l} = \frac{\partial L}{\partial \dot{q}_{l}} = 
\sqrt{2m[E(V_{1}(q_{1}) + V_{2}(q_{2})) - (U_{1}(q_{1}) + 
U_{2}(q_{2}))]}\times 
\frac{\dot{q}_{l}}{\sqrt{ (\dot{q}_{1}^{2} + \dot{q}_{2}^{2})}}
\end{equation}
and obtains a vanishing Hamiltonian, which is a result of 
reparametrization 
invariance,   and a first class constraint $\phi$ as the generator of 
reparametrization gauge symmetry,
\begin{eqnarray}
H &=& p_{l}\dot{q}_{l} - L = 0\nonumber\\
\phi(q_{l}, p_{l}) &=& \frac{1}{V_{1}(q_{1}) + 
V_{2}(q_{2})}\{\frac{1}{2m}
(p_{1}^{2} + p_{2}^{2}) + U_{1}(q_{1}) + U_{2}(q_{2})\} - E \simeq 0
\end{eqnarray}

Following Dirac[8], one may then define a {\em total Hamiltonian}
\begin{eqnarray}
H_{T}& =& H + \alpha (q_{l}, p_{l}) \phi (q_{l}, p_{l})\nonumber\\
     &=& \alpha (q_{l}, p_{l}) \phi (q_{l}, p_{l}) \simeq 0
\end{eqnarray}
where an arbitrary function $\alpha (q_{l}, p_{l})$ specifies a 
choiceof {\em gauge} or a choice of the arbitrary parameter 
$\tau$ in (3.1), which parametrizes the orbit  for a given $E$.
  The quantum theory is defined by( up to an operator ordering)
\begin{equation}
 i\hbar \frac{\partial}{\partial \tau}\psi =  \hat{H}_{T}\psi
\end{equation}
with a physical state condition
\begin{equation}
 \hat{ \alpha} (q_{l}, p_{l})\hat{ \phi} (q_{l}, p_{l})\psi_{phy} = 0
\end{equation}
A choice of the specific gauge  $\alpha (q_{l}, p_{l}) = V_{1}(q_{1}) + 
V_{2}(q_{2})$ gives rise to (2.7) and the choice $\alpha (q_{l}, p_{l}) 
= 1 $ gives the conventional static Schroedinger equation (2.4), since 
$\psi$ appearing in these 
equations are physical states.

The basic dynamical principle involved is thus identified as the 
Jacobi's principle of least 
action, which is analogous to geometrical optics, and the 
formula  of an evolution operator (2.9) dictated by (3.5) 
provides a basis for the path integral approach 
to a general separable Hamiltonian of Liouville-type. The path 
integral in (2.9) deals with a sum over orbits in space instead 
of space-time, and the notion of re-scaled time does not 
explicitly appear in the present approach; the evolution 
operator (2.9) essentially generates a {\em gauge transformation}.

\section{Hydrogen Atom}

It is known that the Schroedinger equation for the hydrogen atom
is written in parabolic coordinates as 
\begin{eqnarray}
\hat{\tilde{H}}_{T}\psi &=& 0\nonumber\\
(\hat{p}_{\varphi} -  \hat{p}_{{\varphi}^{\prime}})\psi& =& 0
\end{eqnarray}
where
\begin{eqnarray}
\hat{\tilde{H}}_{T}&=& \frac{1}{2m}[\hat{p}_{u}^{2} + 
\frac{1}{u^{2}}\hat{p}_{\varphi}^{2} + \hat{p}_{v}^{2} +  
\frac{1}{v^{2}}\hat{p}_{{\varphi}^{\prime}}]  + 
\frac{m\omega^{2}}{2}(u^{2} +v^{2}) - e^{2}\nonumber\\
&=&\frac{1}{2m}\vec{p}_{u}^{2} + \frac{m\omega^{2}}{2}\vec{u}^{2} +
   \frac{1}{2m}\vec{p}_{v}^{2} + \frac{m\omega^{2}}{2}\vec{v}^{2} - 
e^{2}
\end{eqnarray}
and we  defined 
\begin{eqnarray}
\vec{u} &=& ( u_{1},u_{2})= (u\cos\varphi, u\sin\varphi)\nonumber\\
\vec{p}_{u}^{2} &=& \hat{p}_{u}^{2} + 
\frac{1}{u^{2}}\hat{p}_{\varphi}^{2}\nonumber\\
\vec{v} &=& ( v_{1},v_{2})= 
(v\cos{\varphi^{\prime}},v\sin{\varphi^{\prime}})\nonumber\\
\vec{p}_{v}^{2} &=& \hat{p}_{v}^{2} + 
\frac{1}{v^{2}}\hat{p}_{{\varphi}^{\prime}}^{2}
\end{eqnarray} 
The subsidiary condition in (4.1) replaces the use of the
Kustaanheimo-Stiefel transformation[2], and at the same time it 
renders a quadratic Hamiltonian of Liouville-type. This 
introduction of auxiliary variables (4.3) has been discussed by 
Ravndal and Toyoda[5]. 

The general procedure we discussed is thus applicable to the 
hydrogen atom, and we recover the classical exact result of Duru
and Kleinert[1].
\\
\\
{\bf Quantum gravity with a quantized 
cosmological constant}
\\
\\
A alternative way to see the physical meaning of the parameter 
$\tau$ is to study the one-dimensional quantum gravity coupled 
to matter variables $\vec{x}$ defined by
\begin{equation}
\int \frac{{\cal D}\vec{x}{\cal D}h}{gauge\  volume} 
exp\{\frac{i}{\hbar}\int_{0}^{\tau} Lhd\tau\}
\end{equation}
with 
\begin{equation}
L = \frac{m}{2h^{2}}(\frac{d\vec{x}}{d\tau})^{2} - V(r) + E
\end{equation}
where $h$ stands for the einbein, a one-dimensional analogue of 
vierbein $h_{\mu}^{a}$, and $h= \sqrt{g}$ in one-dimension.
If one uses the solution of the equation of motion for $h$ 
defined by
the Lagrangian ${\cal L} = L h$, the action  in (4.4) is reduced 
to the one appearing in the Jacobi's principle of least action.
An interesting aspect of this interpretation is that one has a
toy model of quantum gravity with a quantized cosmological 
constant[6].

\section{Discussion}
An attempt to solve the Green's function for the hydrogen atom exactly 
in the 
path integral[1] opened a 
new avenue for the path integral treatment of a general separable 
Hamiltonian 
of Liouville-type.  This  new view point,
which has been shown to be  based on the Jacobi's principle of least 
action ,  provides a more flexible framework of  path integral to deal 
with a wider class of problems of physical interest. 
The Jacobi's principle of least action , besides being reparametrization 
invariant, gives  a geometrical picture of particle orbits in a 
curved space deformed  by the potential. On the other hand, the 
fundamental 
space-time picture of the conventional Feynman path integral, which is
associated with the Hamilton's principle of stationary action, is lost. 

I happened to learn the path integral of hydrogen atom through a
seminar given by C. Bernido, who provided me references to the 
works of Kleinert. I  wrote 
two papers on this subject, as quoted in this note[4][6]. Though 
I have spent much time on the path integral in relativistic 
quantum field theory, these two papers are the only ones 
I wrote on the path integral of non-relativistic quantum 
mechanics. By publishing these two papers, I was embarassed
to find an enormous difference of {\em culture} between 
physicists in relativistic and 
non-relativistic fields of path integral; 
the major issue in the latter field of non-relativistic path 
integral is 
{\em operator ordering}, which is not fundamental in relativistic 
problems mainly due to the strong constraints provided by 
the symmetry principles such as Lorentz and gauge  invariance.
An over-emphasis on the operator ordering problem does not 
appear to be healthy.

I believe that a more fruitful field will open if the researchers 
in these two realms of path integral work together. 
Professor H. Kleinert will certainly be one of the leaders in 
such an attempt.


\begin{thebibliography}{1}

\bibitem{1}
I. H. Duru and H. Kleinert, Phys. Lett.{\bf B84}(1979)185. 
\bibitem{2}
P. Kustaanheimo and E. Stiefel, J. Reine Angew. Math.{\bf 218}(1965)204.
\bibitem{3}
As for complete references to the related problem, 
see\\ 
H. Kleinert, {\em Path Integrals in Quantum
Mechanics, Statistics and Polymer Physics}(World Scientific, 
Singapore, 1995).
\bibitem{4}
K.Fujikawa, Nucl. Phys.{\bf B484}(1997)495.
\bibitem{5}
F. Ravndal and T. Toyoda, Nucl. Phys. {\bf B3}(1967)312.
\bibitem{6}
K.Fujikawa, Prog. Theor. Phys. 96(1996)863. 
\bibitem{7}
W. Pauli, Ausgewaehlte Kapitel aus der Feldquantisierung, 
Lecture Notes, 
Zurich, 1951.\\
Ph. Choquard, Helv. Phys. Acta {\bf 28}(1955)89.
\bibitem{8}
P. A. M. Dirac, {\em Lectures on Quantum Field Theory}
(Yeshiva Univ., New York, 1966).

\end{thebibliography}
\end{document}